\documentclass{article}

\usepackage{psfig}
\usepackage{rotate}
\usepackage[dvips]{graphicx}

\def\l {\lambda }

\def \ud {{1 \over 2} }

\def \bea {\begin{equation} }
\def \eea {\end{equation} }
\def \cc {coupling constant }

\def \Eslash {E \kern-.5em\slash }
\def \pslash {p \kern-.5em\slash }
\def \kslash {k \kern-.5em\slash }
\def \mij {m_{ij}}
\def \mjk {m_{jk}}
\def \mik {m_{ik}}
\def \mm {m_{3/2}}
\def \mi {m_{\nu_i}}
\def \mj {m_{e_j}}
\def \mk {m_{e_k}}
\def \mips {m_{\nu_i}}
\def \mjps {m_{d_j}}
\def \mkps {m_{d_k}}
\def \mipt {m_{e_i}}
\def \mjpt {m_{u_j}}
\def \mkpt {m_{d_k}}
\def \mipp {m_{u_i}}
\def \mjpp {m_{d_j}}
\def \mkpp {m_{d_k}}
\def \aa {m_{\tilde \nu_{iL}}}
\def \bb {m_{\tilde e_{jL}}}
\def \cc {m_{\tilde e_{kR}}}
\def \aaps {m_{\tilde \nu_{iL}}}
\def \bbps {m_{\tilde d_{jL}}}
\def \ccps {m_{\tilde d_{kR}}}
\def \aapt {m_{\tilde e_{iL}}}
\def \bbpt {m_{\tilde u_{jL}}}
\def \ccpt {m_{\tilde d_{kR}}}
\def \aapp {m_{\tilde u_{iR}}}
\def \bbpp {m_{\tilde d_{jR}}}
\newcommand{\rpv}{\mbox{$\not \hspace{-0.10cm} R_p$ }}
\newcommand{\rpvi}{\mbox{$\not \hspace{-0.10cm} R_p$}}

\if@twoside\oddsidemargin25mm
\else\oddsidemargin25mm\evensidemargin25mm\marginparwidth25mm\fi
\begin{document}

\large

\title{\bf R-parity violation and the cosmological gravitino
problem}
\author{G. Moreau$^1$ and M. Chemtob$^2$ \\ \\
{\it 1: Centre de Physique Th\'eorique, CNRS, Luminy, Case 907 }\\
{\it F-13288 Marseille Cedex 9, France }\\
{\it 2: Service de Physique Th\'eorique, CE-Saclay }\\
{\it F-91191 Gif-sur-Yvette Cedex, France }}
\maketitle

\begin{abstract}
Based on the R-parity violation option of the minimal 
supersymmetric Standard Model, we
exa\-mine the scenario where the massive gravitino, relic from the hot
big-bang, is the lightest supersymmetric particle and can decay through
one or several of the trilinear R-parity violating interactions. We
calculate the rates of the gravitino decay via the various
three-body decay channels with final states involving three quarks
and/or leptons. By taking into account the present constraints on 
the trilinear R-parity violating coupling constants and assuming
the gravitino and scalar superpartner masses do not exceed 
${\cal O}(10TeV)$, it turns out that the gravitinos could easily have
decayed before the present epoch but not earlier than the big-bang 
nucleosynthesis one. Therefore, the considered scenario
would upset the standard big-bang nucleosynthesis and we conclude that it
does not seem to constitute a natural solution for the cosmological gravitino
problem. 
\end{abstract}

\vskip .5cm


\section{Introduction}
\label{intro}

In supergravity theories \cite{Nilles}, the gravitino, namely
the spin-$3/2$ supersymmetric partner of the graviton,
weakly interacts with all the particle species (including itself)
due to the small gravitational strength coupling $\sqrt{G_N}=1/M_P$,
$G_N$ being the gravitational constant and $M_P$ the Planck scale.
Hence, the gravitino-gravitino
annihilation rate is extremely small so that the gravitinos should
decouple at an early epoch of the universe history, and moreover
at an epoch characterized typically by a temperature $T$ higher than
the gravitino mass: $kT>m_{3/2}$ ($k$ being the {\it Boltzmann} constant)
\cite{Wein}. Therefore, the relic abundance of the gravitino should
be large, which is often denoted in the literature as the
cosmological ``gravitino problem''. One of the first solutions for the 
gravitino problem to be envisaged is to compensate the
large gravitino relic abundance by a gravitino mass
sufficiently small, namely $m_{3/2}<1 keV$ \cite{Pagels}, to
respect the limit on the present universe energy density: $\Omega_0
\stackrel{<}{\sim} 1$.
For heavier gravitinos, a second type of available solution is by
shortening their lifetime so that they do not survive out at the late
epochs \cite{Wein}.
This can be realized in two characteristic options:
either the gravitino is not
the Lightest Supersymmetric Particle (LSP) and thus can decay into
an odd number of superpartners through gravitational and gauge
interactions (both of these ones couple an even number of superpartners), 
or it is the LSP. A gravitino LSP can be realized within
various supersymmetric models, including the gauge mediated supersymmetry
breaking models \cite{Dine1,Dine2}, the models of low fundamental energy
scale \cite{Arkani} and even the conventional gravity mediated supersymmetry
breaking models (for some specific set of the supersymmetry breaking
parameters). If the gravitino is the LSP, it can decay
only into the ordinary particles of the Standard Model. Such a decay
channel must involve \cite{rev82} both gravitational and the so-called
R-parity
symmetry \cite{Salam,Fayet1} violating interactions (the latter ones couple
an odd
number of superpartners). The R-parity violating (\rpvi)
interactions are written in the following superpotential, in terms of the
left-handed superfields for the leptons ($L$), quarks ($Q$) and Higgs of
hypercharge $1/2$ ($H$) and the right-handed superfields for the charged
leptons ($E^c$), up and down type quarks ($U^c,D^c$),
\begin{eqnarray}
W_{\rpv}=\sum_{i,j,k} \bigg (\ud \l _{ijk} L_iL_j E^c_k+
\l^{\prime}_{ijk} L_i Q_j D^c_k+ \ud \l^{\prime \prime}_{ijk}
U_i^cD_j^cD_k^c
+ \mu_i H L_i \bigg ),
\label{superpot}
\end{eqnarray}
$i,j,k$ being flavor indices,
$\l_{ijk},\l^{\prime}_{ijk},\l^{\prime \prime}_{ijk}$
dimensionless coupling constants and $\mu_i$ dimension one parameters.
Note that in the scenario in which the gravitino is not the LSP,
the gravitino preferentially decays into an ordinary Standard Model particle
and its superpartner through
gravitational interactions, since the \rpv coupling constants
are severely constrained by the low-energy experimental bounds obtained at
colliders \cite{Drein,Bhatt}.

In the scenario of an unstable gravitino heavier than the LSP, the
gravitino decay can produce an unacceptable amount of LSP which conflicts with
the observations of the present mass density of the universe \cite{Krauss}.
The scenario containing an unstable LSP gravitino, having decay channels
which involve \rpv coupling constants, is based on the violation of
the R-parity symmetry. Now, neither the grand unified theories,
the string theories nor the study of the discrete gauge symmetries give
a strong theoretical argument in favor of the conservation of the R-parity
symmetry in the supersymmetric extension of the Standard Model \cite{Drein}.
Hence, the scenario with an unstable LSP gravitino constitutes an attractive
possibility which must be considered as an original potential solution
with respect to the cosmological gravitino problem.

Our main purpose in the present work is to determine whether the
scenario of an un\-stable LSP gra\-vitino decaying via \rpv interactions
constitutes effectively a natural solution to the cosmological 
gravitino problem. We will concentrate on the trilinear \rpv interactions,
namely $\l _{ijk} L_iL_j E^c_k$, $\l^{\prime}_{ijk} L_i Q_j D^c_k$ and
$\l^{\prime \prime}_{ijk} U_i^cD_j^cD_k^c$. With the goal of
enhancing the gra\-vitino instability, we will consider an optimistic type
of scenario in which several trilinear \rpv coupling constants have
simultaneously non-vanishing values. We will not consider the bilinear
\rpv term $\mu_i H L_i$ of Eq.(\ref{superpot}) which can always be rotated
away by a suitable redefinition of superfields \cite{redef}.
The bilinear \rpv interactions as well as the pos\-sible
alternative of a spontaneous breaking of the R-parity symmetry have
been considered, within the context of the cosmological gravitino
problem, in a recent study \cite{Taka} which examines the two-body 
gravitino decay mode into photon and neutrino (we will compare the results
of \cite{Taka} with ours). We also mention here the works \cite{Lee1}
and \cite{Lee2} in which have been considered the \rpv decay reactions 
of single nucleon into a gravitino and a strange meson.

In Section \ref{sec:cons}, we determine the constraints arising from
the cosmological gravitino pro\-blem in a scenario characterized by an
unstable gravitino. In Section \ref{sec:rate}, we calculate the dominant
gravitino decay rates involving the trilinear \rpv interactions.
Finally, in Section \ref{sec:disc}, we give a
quantitative discussion aimed at determining whether the scenario of an
unstable LSP gravitino decaying through \rpv interactions can effectively
constitute a natural solution with regard to the cosmological gravitino
problem. This quantitative discussion is based on the results of
Sections \ref{sec:cons} and \ref{sec:rate} as well as the present
low-energy constraints on the \rpv coupling constants \cite{Bhatt}.

\section{Constraints on the gravitino decay temperature}
\label{sec:cons}

In this section, we derive the constraints \cite{Wein} characteristic of a
scenario
containing an unstable gravitino supposed to solve the cosmological
gravitino problem.

At the temperature $kT \approx m_{3/2}$, the expansion rate of the universe
is $H(kT \approx m_{3/2}) \approx
\sqrt{m_{3/2}}(kT)^{3/2}/M_P=m_{3/2}^2/M_P$.
Besides, the gravitino decay having the larger rate is the two-body decay
into a particle of the Standard Model and its supersymmetric partner. The
largest gravitino decay rate is thus of order,
\begin{eqnarray}
\Gamma_{max} \sim m_{3/2}^3/M_P^2,
\label{max}
\end{eqnarray}
which is smaller than $H(kT \approx m_{3/2})$ by a factor $M_P/m_{3/2}$.
Therefore, the equilibrium condition,
\begin{eqnarray}
\Gamma \geq H(T),
\label{equi}
\end{eqnarray}
$\Gamma$ being the gravitino decay rate,
can only be reached for temperatures much smaller
than $kT \approx m_{3/2}$. In other words, the gravitinos cannot decay as
long as the temperature has not dropped far below the gravitino mass.
This conclusion has two important consequences.

The first consequence is that the gravitino decays occur at temperatures
at which the gravitinos cannot be created in collisions, so that the
gravitino population can effectively be reduced by the gravitino decays.

The other consequence is that the equilibrium condition (Eq.(\ref{equi}))
is reached at an epoch at which the cosmic energy density is dominated by
the gravitino energy density $\rho_{3/2}$.
Therefore, in determining the temperature $T_{3/2}$ at which the
equilibrium condition of Eq.(\ref{equi}) is reached (the gravitino decay
temperature) by solving the equation $\Gamma = H(T_{3/2})$, one must
take the following expression for the expansion rate,
\begin{eqnarray}
H(T) \approx \sqrt{{8 \pi G_N \over 3} \rho_{3/2}}.
\label{exp1}
\end{eqnarray}
Since the gravitino mass density reads as,
\begin{eqnarray}
\rho_{3/2} = {3 \zeta(3) \over \pi^2} {g(T) \over g(T_d)} m_{3/2} (kT)^3,
\label{dens}
\end{eqnarray}
where $\zeta (x)$ is the {\it Riemann} zeta function 
and $\zeta(3)=1.20206\dots$,
the equation Eq.(\ref{exp1}) can be rewritten as,
\begin{eqnarray}
H(T) \approx  \sqrt{{8 \zeta(3) \over \pi}{g(T) \over g(T_d)}}
{\sqrt{m_{3/2}} \over M_P} (kT)^{3/2}.
\label{exp2}
\end{eqnarray}
In Eq.(\ref{dens}) and Eq.(\ref{exp2}), g(T) counts the total number
of effectively massless degrees of freedom (those of species with mass
$m \ll kT$), and $T_d$ is the gravitino decoupling temperature, namely
the temperature at which is reached the out of equilibrium condition
$\Gamma_{an} < H(T)$, $\Gamma_{an}$ being the gravitino-gravitino
annihilation rate. The gravitino decoupling temperature is typically
larger than the gravitino mass, as we have mentioned in the introduction.
The presence of the factor $g(T)/g(T_d)$ in Eq.(\ref{dens})
and Eq.(\ref{exp2}) allows to take into account the increase of the
temperature due to the conservation of the entropy per comoving volume
$S=(2 \pi^2 / 45) g(T) (kT)^3 R^3$, where $R=R(t)$ is the time dependent 
scale factor. From Eq.(\ref{exp2}) it follows that the gravitino
decay temperature is given by,
\begin{eqnarray}
kT_{3/2} \approx \bigg ({\pi \over 8 \zeta(3)}{g(T_d) \over g(T_{3/2})}\bigg )^{1/3}
\bigg ({\Gamma^2 M_P^2 \over m_{3/2}}\bigg )^{1/3}.
\label{Tc}
\end{eqnarray}
At the temperature $T_{3/2}$, most of the gravitinos decay. After the
gravitinos have decayed and their decay energy has been thermalized,
the temperature rises to the value $kT^\prime_{3/2}$ given by,
\begin{eqnarray}
kT^\prime_{3/2} \approx {(90 \zeta(3))^{1/4} \over \pi}
\bigg ({m_{3/2} (kT_{3/2})^3 \over g(T_{3/2})}\bigg )^{1/4}.
\label{Tcp1}
\end{eqnarray}
The equation Eq.(\ref{Tcp1}) is nothing else but the expression of
the energy density conservation. From Eq.(\ref{Tc}) and Eq.(\ref{Tcp1})
it follows that $kT^\prime_{3/2}$ reads as,
\begin{eqnarray}
kT^\prime_{3/2} \approx  \bigg ( {45 \over 4 \pi^3} \bigg )^{1/4}
{g(T_d)^{1/4} \over g(T_{3/2})^{1/2}} \sqrt{\Gamma M_P}.
\label{Tcp2}
\end{eqnarray}

In order to avoid large relic abundance of an unstable gravitino, this
one must decay before the present epoch. This requirement imposes
the following bound on the increased gravitino decay temperature
$T^\prime_{3/2}$,
\begin{eqnarray}
T^\prime_{3/2} > 2.75 K \ \ (kT^\prime_{3/2} > 2.36 \ 10^{-10} MeV).
\label{bound1}
\end{eqnarray}

Furthermore, the decay of the gravitinos leads to an increase of the
temperature by a factor which can be deduced from Eq.(\ref{Tc}) and
Eq.(\ref{Tcp2}):
\begin{eqnarray}
{T^\prime_{3/2} \over T_{3/2}}
& \approx & {45^{1/4} \sqrt{2} \zeta(3)^{1/3} \over \pi^{13/12}}
\bigg (g(T_{3/2})\sqrt{g(T_d)}\bigg )^{-1/6} \bigg ({m_{3/2}
\over \sqrt{\Gamma M_P}} \bigg )^{1/3} \cr
& > & {45^{1/4} \sqrt{2} \zeta(3)^{1/3} \over \pi^{13/12}}
\bigg (g(T_{3/2})\sqrt{g(T_d)}\bigg )^{-1/6} \bigg ({M_P
\over m_{3/2}} \bigg )^{1/6}.
\label{Tcinc}
\end{eqnarray}
In order to derive the bound of Eq.(\ref{Tcinc}), we have used the fact
that $\Gamma < \Gamma_{max}$ (see Eq.(\ref{max})). This increase of the
temperature leads to an increase of 
the entropy density $s=(2 \pi^2 / 45) g(T) (kT)^3$
and hence to a decrease of the baryon-to-entropy ratio (or baryon number
in a comoving volume) $B=n_B/s$, $n_B$ being the baryon number density.
Therefore, if the gravitinos decay after the nucleosynthesis epoch, the
baryon-to-entropy ratio during the nucleosynthesis epoch must be much
greater than the present one, leading to the production
through nucleosynthesis of too much helium
and too little deuterium (compared with the constraints on the primordial
abundances derived from observational data) \cite{Wein}. In conclusion,
if the gravitinos decay before the present epoch, those decays have to
occur before the nucleosynthesis epoch, or in other
words, the increased gravitino decay temperature $T^\prime_{3/2}$ has to be
higher than the nucleosynthesis temperature, namely,
\begin{eqnarray}
k T^\prime_{3/2} > 0.4 MeV.
\label{bound2}
\end{eqnarray}

In summary, in a scenario of unstable gravitino protected against large
gravitino relic abundance, the gravitino must decay before the present
epoch,
namely $kT^\prime_{3/2} > 2.36 \ 10^{-10} MeV$ (see Eq.(\ref{bound1})), or even
at
an earlier epoch than the nucleosynthesis one, namely $kT^\prime_{3/2} > 0.4
MeV$
(see Eq.(\ref{bound2})), if one requires that this scenario does not
conflict
with the current cosmological ideas on nucleosynthesis. Hence, in order to
determine whether a given scenario of unstable
gravitino can effectively constitute a solution to the gravitino
problem, one has to calculate, within this scenario, the gravitino decay
rate.
Indeed, the value of the gravitino decay rate gives, through
Eq.(\ref{Tcp2}),
the gravitino decay temperature $T^\prime_{3/2}$, and thus allows to verify
whether the constraints of Eq.(\ref{bound1}) and Eq.(\ref{bound2}) are
respected or not. The calculation of the gravitino decay rate will be done
in the next section, within the scenario we consider.

\section{Rates of the gravitino decay via trilinear
R-parity violating interactions}
\label{sec:rate}

\begin{figure}
\centerline{\psfig{figure=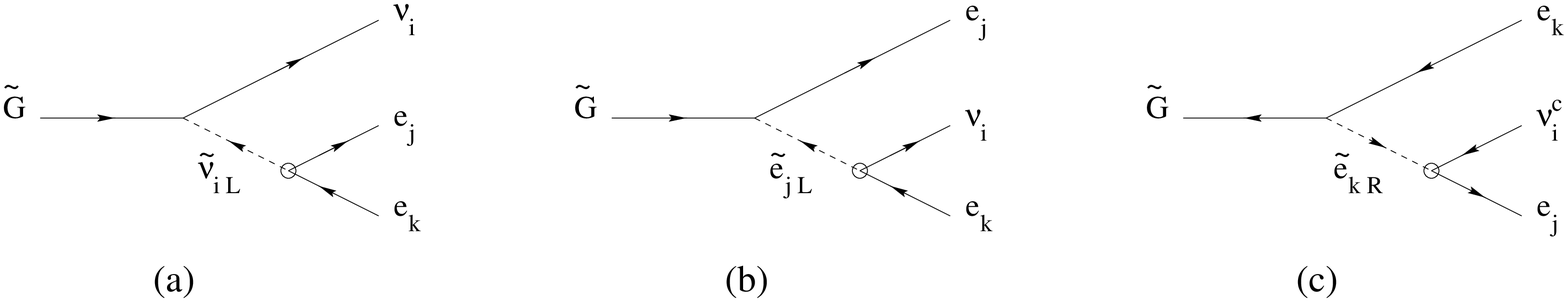,height=5.cm,width=18.cm}}
\caption{{\it Feynman} diagrams of the gravitino three-body decay processes
involving
the $\l_{ijk}L_iL_jE_k^c$ interactions. These interactions
are represented by the circled vertex. The other vertex correspond
to gravitational couplings. The plain (dashed) lines symbolize the
propagation
of fermionic (scalar) particles. The arrows denote the flow of momentum of
the associated particles. Finally, $e$, $\tilde e$, $\nu$, $\tilde \nu$
and $\tilde G$ represent respectively the charged leptons, the charged
sleptons, the neutrinos, the sneutrinos and the gravitino, $i,j,k$
are generation indices, the $L, R$ indices stand for Left, Right
(chirality) and the $c$ exponent indicates a charge conjugated particle.
We have not drawn the charge conjugated reactions which can be
trivially deduced from these ones.}
\label{grdec1}
\end{figure}

\begin{figure}
\centerline{\psfig{figure=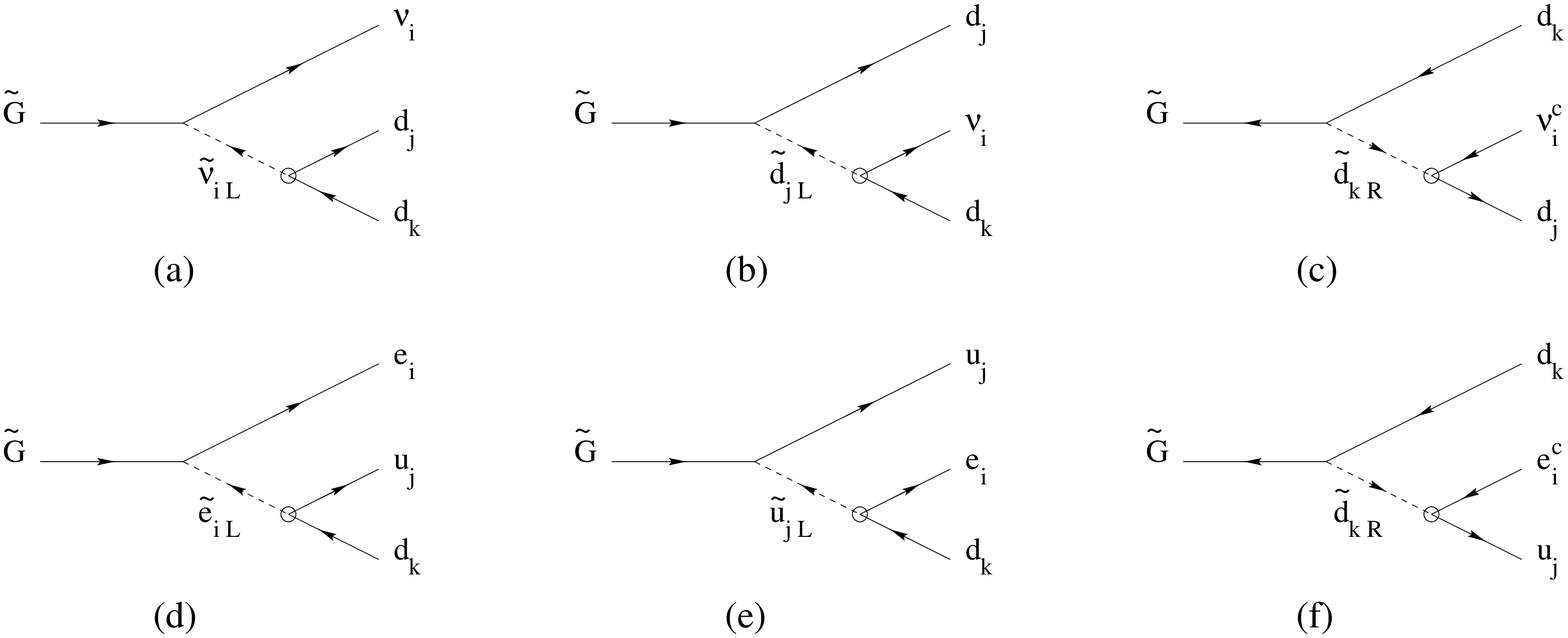,height=10.cm,width=18.cm}}
\caption{{\it Feynman} diagrams of the gravitino three-body decay processes
involving
the $\l^{\prime}_{ijk}L_iQ_jD_k^c$ interactions. These interactions
are represented by the circled vertex. The other vertex correspond
to gravitational couplings. The plain (dashed) lines symbolize the
propagation
of fermionic (scalar) particles. The arrows denote the flow of momentum of
the associated particles. Finally, $e$, $\tilde e$, $\nu$, $\tilde \nu$,
$d$, $\tilde d$, $u$, $\tilde u$ and $\tilde G$ represent respectively the
charged leptons, the charged sleptons, the neutrinos, the sneutrinos, the
down quarks, the down squarks, the up quarks, the up squarks and the
gravitino, $i,j,k$ are generation indices, the $L, R$ indices stand for
Left, Right (chirality) and the $c$ exponent indicates a charge conjugated
particle. We have not drawn the charge conjugated reactions which can be
trivially deduced from these ones.}
\label{grdec2}
\end{figure}

\begin{figure}
\centerline{\psfig{figure=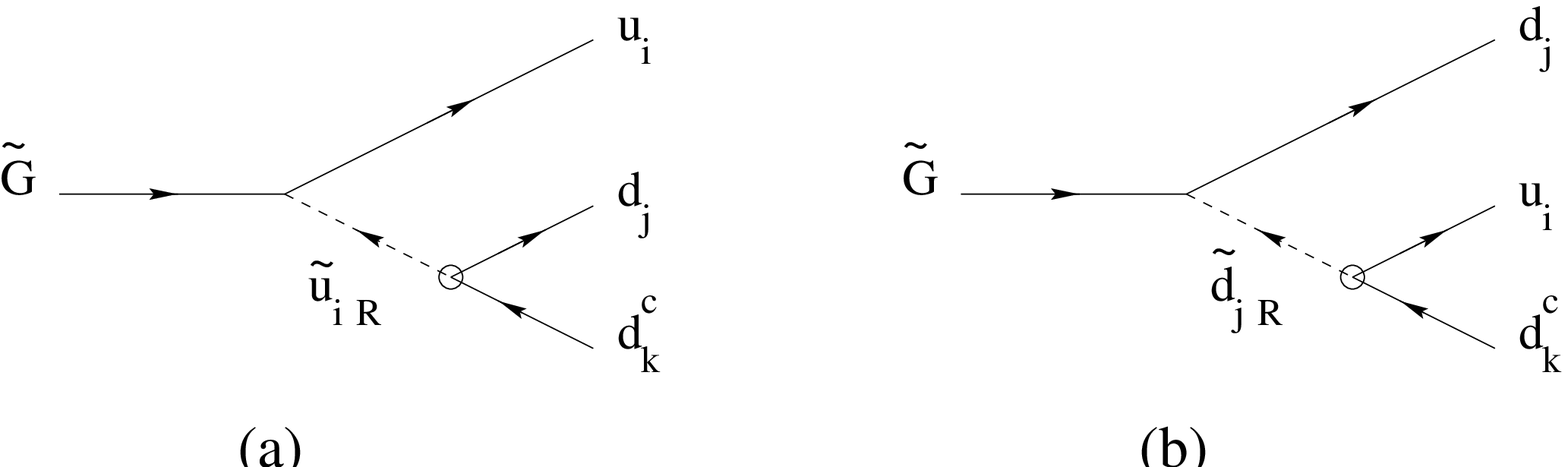,height=5.cm,width=12.cm}}
\caption{{\it Feynman} diagrams of the gravitino three-body decay processes
involving
the $\l^{\prime \prime}_{ijk}U_i^cD_j^cD_k^c$ interactions. These
interactions
are represented by the circled vertex. The other vertex correspond
to gravitational couplings. The plain (dashed) lines symbolize the
propagation
of fermionic (scalar) particles. The arrows denote the flow of momentum of
the associated particles. Finally, $d$, $\tilde d$, $u$, $\tilde u$ and
$\tilde G$ represent respectively the down quarks, the down squarks, the up
quarks, the up squarks and the gravitino, $i,j,k$ are generation indices,
the $L, R$ indices stand for Left, Right (chirality) and the $c$ exponent
indicates a charge conjugated particle. We have not drawn the charge
conjugated
reactions which can be trivially deduced from these ones.}
\label{grdec3}
\end{figure}

As we have discussed in the introduction, in the scenario we study, the
gravitino is the LSP, so that it can decay only via \rpv interactions, and
the R-parity symmetry is violated only by the trilinear terms of the
superpotential \ref{superpot}. Hence, in our scenario, the gravitino
can only decay through the trilinear \rpv couplings of Eq.(\ref{superpot}).
The gravitino decay processes which involve the trilinear \rpv interactions
of Eq.(\ref{superpot}) and have the dominant rates are obviously the decays
through a virtual scalar superpartner into three ordinary fermions involving
one gravitational and one \rpv coupling. Those processes are represented in
Fig.\ref{grdec1}, Fig.\ref{grdec2} and Fig.\ref{grdec3}
for the \rpv interactions of type $\l_{ijk}L_iL_jE_k^c$,
$\l^{\prime}_{ijk}L_iQ_jD_k^c$ and $\l^{\prime \prime}_{ijk}U_i^cD_j^cD_k^c$
(see Eq.(\ref{superpot})), respectively. The {\it Feynman} diagrams of
Fig.\ref{grdec1}, Fig.\ref{grdec2} and Fig.\ref{grdec3} have been derived
from the structures of the relevant interaction lagrangians. Let us collect
these relevant interaction lagrangians. First, the locally supersymmetric
lagrangian term which describes the coupling between the fields of the
gravitino
($\Psi_\mu$), an ordinary fermion ($\psi$) and its scalar superpartner
($\phi$)
reads as,
\begin{eqnarray}
{\cal L}=-{1 \over \sqrt{2} M_\star} \bar \psi_L \gamma^\mu \gamma^\nu
\partial_\nu \phi \Psi_{\mu R} \ + \ \mbox{h.c.},
\label{lagrav}
\end{eqnarray}
the $L$/$R$ indices standing for Left/Right chirality, $\gamma^\mu$ being
the Dirac
matrices and $M_\star=(8 \pi G_N)^{-1/2}=2.4 \ 10^{18} GeV$ the reduced
Planck mass.
With our conventions, the reduced Planck mass $M_\star$ and the gravitino
mass
$m_{3/2}$ are related through the formula,
\begin{eqnarray}
m_{3/2}={F \over \sqrt{3} M_\star},
\label{lagrel}
\end{eqnarray}
where $F$ denotes the scale of spontaneous supersymmetry breaking.
Secondly, the relevant lagrangian of the \rpv interactions of type $\l$,
$\l^{\prime}$ and $\l^{\prime \prime}$, originating from the superpotential
\ref{superpot}, is the following,
\begin{eqnarray}
{\cal L}_{\rpv}= \sum_{ijk} \bigg [ -\lambda_{ijk} && {1 \over 2}
\bigg ( \tilde  \nu_{iL}\bar e_{kR}e_{jL} +
\tilde e_{jL}\bar e_{kR}\nu_{iL} + \tilde e^\star _{kR}\bar \nu^c_{iR}
e_{jL}-(i \leftrightarrow j) \bigg ) \cr
- \lambda^{\prime}_{ijk} && \bigg ( \tilde  \nu_{iL}\bar d_{kR}d_{jL} +
\tilde d_{jL}\bar d_{kR}\nu_{iL} + \tilde d^\star _{kR}\bar \nu^c_{iR}
d_{jL} \cr && -\tilde  e_{iL}\bar d_{kR}u_{jL} -
\tilde u_{jL}\bar d_{kR}e_{iL} - \tilde d^\star _{kR}\bar e^c_{iR} u_{jL}
\bigg ) \cr - \lambda^{\prime \prime}_{ijk} && {1 \over 2}
\bigg (\tilde  u^\star _{i R}\bar d_{j R}d^c_{k L} +
2 \tilde  d^\star _{j R}\bar u_{i R}d^c_{k L} \bigg ) \bigg ]
\ + \ \mbox{h.c.},
\label{lagrpv}
\end{eqnarray}
where $e$, $\nu$, $d$ and $u$ represent respectively the fields of the
charged
leptons, the neutrinos, the down and the up quarks, a \ $\tilde { }$
\ indicates the
scalar superpartner and our notation is such that one has, for instance,
$\bar \nu^c_{iR}=\overline{(\nu^c_i)_R}$.

As shown by Fig.\ref{grdec1}, when the gravitino decays via a $\l_{ijk}$
coupling, it decays into two charged leptons and a neutrino as
$\tilde G \stackrel{\l_{ijk}}{\to} \nu_i e_j \bar e_k,
\bar \nu_i \bar e_j e_k$. Based on the
lagrangians of Eq.(\ref{lagrav}) and Eq.(\ref{lagrpv}), we have
calculated the rate of this gravitino decay reaction, which receives
three contributions as can be observed in Fig.\ref{grdec1}. The full
analytical result for the spin summed amplitude of this gravitino decay
reaction is given in Appendix \ref{formulas}. We give here the
analytical result for the integrated rate of the gravitino decay
$\tilde G \stackrel{\l_{ijk}}{\to} \nu_i e_j \bar e_k$, under the
assumption that the masses of the final state particles are negligible
with respect to the gravitino and exchanged scalar superpartner masses
(we will comment on this hypothesis below),
\begin{eqnarray}
\Gamma(\tilde G \stackrel{\l_{ijk}}{\to} \nu_i e_j \bar e_k) & = &
{1 \over 96 (2 \pi)^3 } {\l_{ijk}^2 \over \mm^3 M_\star^2}
\bigg [ \Omega(\aa)+\Omega(\bb)+\Omega(\cc)
+\Xi(\aa,\cc) \cr && +\Xi(\bb,\cc)+\Xi(\aa,\bb)
+\Sigma(\aa,\cc)+\Sigma(\bb,\cc) \cr && +\Sigma(\aa,\bb) +\Sigma(\bb,\aa)
+  \Delta(\aa,\cc) \cr && [Sp({\aa^2 \over \aa^2 +\cc^2-\mm^2 })-Sp({\aa^2
-\mm^2
\over \aa^2 +\cc^2-\mm^2 })]
+\Delta(\bb,\cc) \cr && [Sp({\bb^2 \over \bb^2 +\cc^2-\mm^2 })-Sp({\bb^2
-\mm^2
\over \bb^2 +\cc^2-\mm^2 })]
+\Delta(\aa,\bb) \cr && [-Sp(-{\bb^2 \over \aa^2})-Sp(-{\aa^2 \over \bb^2})
+Sp({\mm^2 -\bb^2 \over \aa^2})+Sp({\mm^2 -\aa^2 \over
\bb^2})] \bigg ], \cr &&
\label{gamma}
\end{eqnarray}
where the mass functions $\Omega(m)$, $\Xi(m_1,m_2)$, $\Delta(m_1,m_2)$ and
$\Sigma(m_1,m_2)$ are defined in Appendix \ref{definitions},
$Sp(x)=Polylog(2,x)=Li_2(x)$ is the dilogarithm or {\it Spence} function
and $\aa$, $\bb$ and $\cc$ are respectively the masses of the
superpartners $\tilde \nu_{iL}$, $\tilde e_{jL}$ and $\tilde e_{kR}$.
\\ The hypothesis that the masses of the final state particles, in the
reaction $\tilde G \stackrel{\l_{ijk}}{\to} \nu_i e_j \bar e_k$, are
negligible with respect to the gravitino and exchanged scalar superpartner
masses has been made in order to obtain a more simple form for the
analytical
result of the integrated gravitino decay rate (see Eq.(\ref{gamma})).
Nevertheless, in Section \ref{sec:disc}, the numerical results will take
into account the full mass effect of the final state particles.
Moreover, this assumption that the masses of the final state particles
are negligible with respect to the gravitino and exchanged scalar
superpartner
masses represents a good approximation justified by the two following
points.
First, the present experimental limits on the scalar superpartner masses
(noted generically by $\tilde m$), obtained at colliders within an \rpv
model containing a non-vanishing Yukawa coupling of type $\l$, $\l^{\prime}$
or $\l^{\prime \prime}$, are typically of order $\tilde m \stackrel{>}{\sim}
100GeV$ \cite{exp}. Secondly, as we will see, in the quantitative discussion
of Section \ref{sec:disc}, we will not need to consider gravitino masses
smaller than $50GeV$.
\\ Finally, we note that the integrated rate of the gravitino decay
$\tilde G \stackrel{\l_{ijk}}{\to} \bar \nu_i \bar e_j e_k$ is equal to the
integrated rate of the charge conjugated decay $\tilde G
\stackrel{\l_{ijk}}{\to} \nu_i e_j \bar e_k$, which is given by the
formula of Eq.(\ref{gamma}).

The two kinds of gravitino decay channel involving $\l^\prime_{ijk}$
couplings are $\tilde G \stackrel{\l^\prime_{ijk}}{\to} \nu_i d_j \bar d_k,
\bar \nu_i \bar d_j d_k$ and $\tilde G \stackrel{\l^\prime_{ijk}}{\to}
e_i u_j \bar d_k, \bar e_i \bar u_j d_k$ (see Fig.\ref{grdec2}).
Since the structures of the $\l_{ijk}$ and $\l^\prime_{ijk}$ interaction
lagrangians are similar (see Eq.(\ref{lagrpv}), Fig.\ref{grdec1} and
Fig.\ref{grdec2}), the integrated rate of the gravitino decay
 $\tilde G \stackrel{\l^\prime_{ijk}}{\to} \nu_i d_j \bar d_k$
($\tilde G \stackrel{\l^\prime_{ijk}}{\to}   e_i u_j \bar d_k$) is
obtained from the formula of Eq.(\ref{gamma}) simply by changing
$\l_{ijk}$ to $\l^\prime_{ijk}$, adding an extra color factor of $N_c=3$
and replacing the masses as follows: $\aa \to \aaps \ (\aapt)$,
$\bb \to \bbps \ (\bbpt)$ and $\cc \to \ccps \ (\ccpt)$. The approximation
of neglecting the masses of the final state particles is also well
motivated in the calculation of the gravitino decay rate involving
$\l^\prime_{ijk}$ coupling constants, except when the gravitino decays into
the top quark as $\tilde G \stackrel{\l^\prime_{i3k}}{\to} e_i t \bar d_k,
\bar e_i \bar t d_k$. However, recall that in Section \ref{sec:disc}, the
mass effect of the final state particles will be included in the
numerical computations.

In case the gravitino decays through a coupling of type
$\l^{\prime \prime}_{ijk}$, it decays into three quarks as
$\tilde G \stackrel{\l^{\prime \prime}_{ijk}}{\to} u_i d_j d_k,
\bar u_i \bar d_j \bar d_k$, as illustrates Fig.\ref{grdec3}.
In Appendix \ref{formulas}, we give the full analytical result
for the spin summed amplitude of this gravitino decay process.
In the following, we present the analytical result for the integrated
rate of the gravitino decay $\tilde G
\stackrel{\l^{\prime \prime}_{ijk}}{\to} u_i d_j d_k$, in the
hypothesis the masses of the final state particles are negligible
compared with the gravitino and exchanged scalar superpartner
masses (see above for detailed comments on this approximation),
\begin{eqnarray}
\Gamma(\tilde G \stackrel{\l^{\prime \prime}_{ijk}}{\to} u_i d_j d_k) & = &
{N_c ! \over 96 (2 \pi)^3 } {\l^{\prime \prime 2}_{ijk} \over \mm^3
M_\star^2}
\bigg [ \Omega(\aapp)+4 \Omega(\bbpp)+2 \Xi(\aapp,\bbpp) \cr &&
+ 2 \Sigma(\aapp,\bbpp) +2 \Sigma(\bbpp,\aapp)
+ 2 \Delta(\aapp,\bbpp) \cr && [-Sp(-{\bbpp^2 \over \aapp^2})
-Sp(-{\aapp^2 \over \bbpp^2})
+Sp({\mm^2 -\bbpp^2 \over \aapp^2})
+Sp({\mm^2 -\aapp^2 \over \bbpp^2})]
\bigg ], \cr &&
\label{gammapp}
\end{eqnarray}
where $N_c=3$ is the number of colors and
$\Omega(m)$, $\Xi(m_1,m_2)$, $\Delta(m_1,m_2)$ and $\Sigma(m_1,m_2)$
are defined in Appendix \ref{definitions}.

\section{Numerical results and discussion}
\label{sec:disc}

In this section, we present the numerical results for the gravitino
decay temperature $T^\prime_{3/2}$, within the scenario characterized
by an LSP gravitino having a decay channel which involves trilinear
\rpv interactions. Then we will determine whether the values
found for the temperature $T^\prime_{3/2}$ respect or not the limits
given in Eq.(\ref{bound1}) and Eq.(\ref{bound2}). We will thus be
able to conclude on the possibility for the scenario with an unstable
LSP gravitino, decaying via $\l$, $\l^\prime$ or $\l^{\prime \prime}$
couplings, to solve the cosmological gravitino problem.

\vskip 0.4 cm {\bf $\bullet \ \lambda_{ijk}$ couplings} \vskip 0.4 cm 
We first consider the case where one \rpv coupling constant of type
$\l_{ijk}$ is dominant so that all the other \rpv coupling
constants are negligible in comparison. The assumption of a single
dominant \rpv coupling constant is often adopted in the literature
(as we will remark below) for simplification reasons. This assumption
find its justification in the analogy between the structures of the
\rpv and Yukawa couplings, the later ones exhibiting a strong hierarchy
in flavor space. In this first case, the gravitino decay temperature
$T^\prime_{3/2}$ is derived (through Eq.(\ref{Tcp2})) from the gravitino
decay rate involving $\l_{ijk}$ coupling constants $\Gamma(\tilde G
\stackrel{\l_{ijk}}{\to} \nu_i e_j \bar e_k)$, which is given
in Eq.(\ref{gamma}).
\\ In Fig.\ref{astrodec}(a) and Fig.\ref{astrodec}(b),
we present the gravitino decay temperature $T^\prime_{3/2}$ as a
function of the scalar superpartner mass $\tilde m$ for two
gravitino masses, in the case of a single dominant \rpv coupling
constant of type $\l_{ijk}$. Let us detail some particular points
concerning the derivation of those numerical results. First, for
simplification reason, we have set
all the scalar superpartner masses to a common value, denoted by
$\tilde m$. In order to maximize the gravitino decay rate, and thus
the temperature $T^\prime_{3/2}$ (see Eq.(\ref{Tcp2})), we have also
assumed that the single dominant \rpv coupling constant is $\l_{233}$,
which is one of the $\l_{ijk}$ coupling constants having the weakest
present low-energy constraint, namely $\l_{233}<0.06(m_{\tilde \tau_R}
/100GeV)$ \cite{Drein,Barger}, and that $\l_{233}$ is equal to its
present low-energy bound. We note that the present low-energy constraint
on $\l_{233}$ has been obtained under the hypothesis that $\l_{233}$ is
the single dominant \rpv coupling constant \cite{Barger}, which is
consistent with our assumptions. At this stage, a particularity due the
structure of the $\l_{ijk}$ couplings must be described. Since the
$\l_{ijk}$ coupling constants are anti-symmetric in the $i$ and $j$
family indices, in the considered case where $\l_{233}$ is
a dominant \rpv coupling cons\-tant,
$\l_{323}$ is a second dominant \rpv coupling constant (ha\-ving the
opposite value) so that the main gravitino decay channels are $\tilde G
\stackrel{\l_{233}}{\to} \nu_\mu \tau \bar \tau, \bar \nu_\mu \bar
\tau \tau$ and $\tilde G \stackrel{\l_{323}}{\to} \nu_\tau \mu \bar
\tau, \bar \nu_\tau \bar \mu \tau$. This point has been taken into
account in the calculation. The decay temperature in
Eq.(\ref{Tcp2}) is proportional to the combination of statistical
factors $g(T_d)^{1/4}/g(T_{3/2})^{1/2}$. Let us here assume 
tentatively that the thermalized degrees of fredom at the 
gravitino decoupling and decay temperatures are the same as 
those for the minimal supersymmetric Standard Model and 
the present epochs, respectively. U\-sing then the
values $g(T_d) \simeq 915 / 4  \simeq 228.75,\ g(T_{3/2}) =
43 / 11= 3.909$, one obtains: $T^\prime_{3/2} \propto
g(T_d)^{1/4}/g(T_{3/2})^{1/2} \simeq 1.96$. The actual numerical 
value of this factor might be larger but not by very much. In the numerical
results, the above ratio of statistical factors has been set to unity
(we will come back to this point later). 
Finally, the gravitino decay rate has
been multiplied by a factor of $2$ in order to count the charge
conjugated gravitino decay process and the masses of the final state
particles have been taken into account in the computation.
\begin{figure}
\begin{center}
\leavevmode
\includegraphics[height=4.cm]{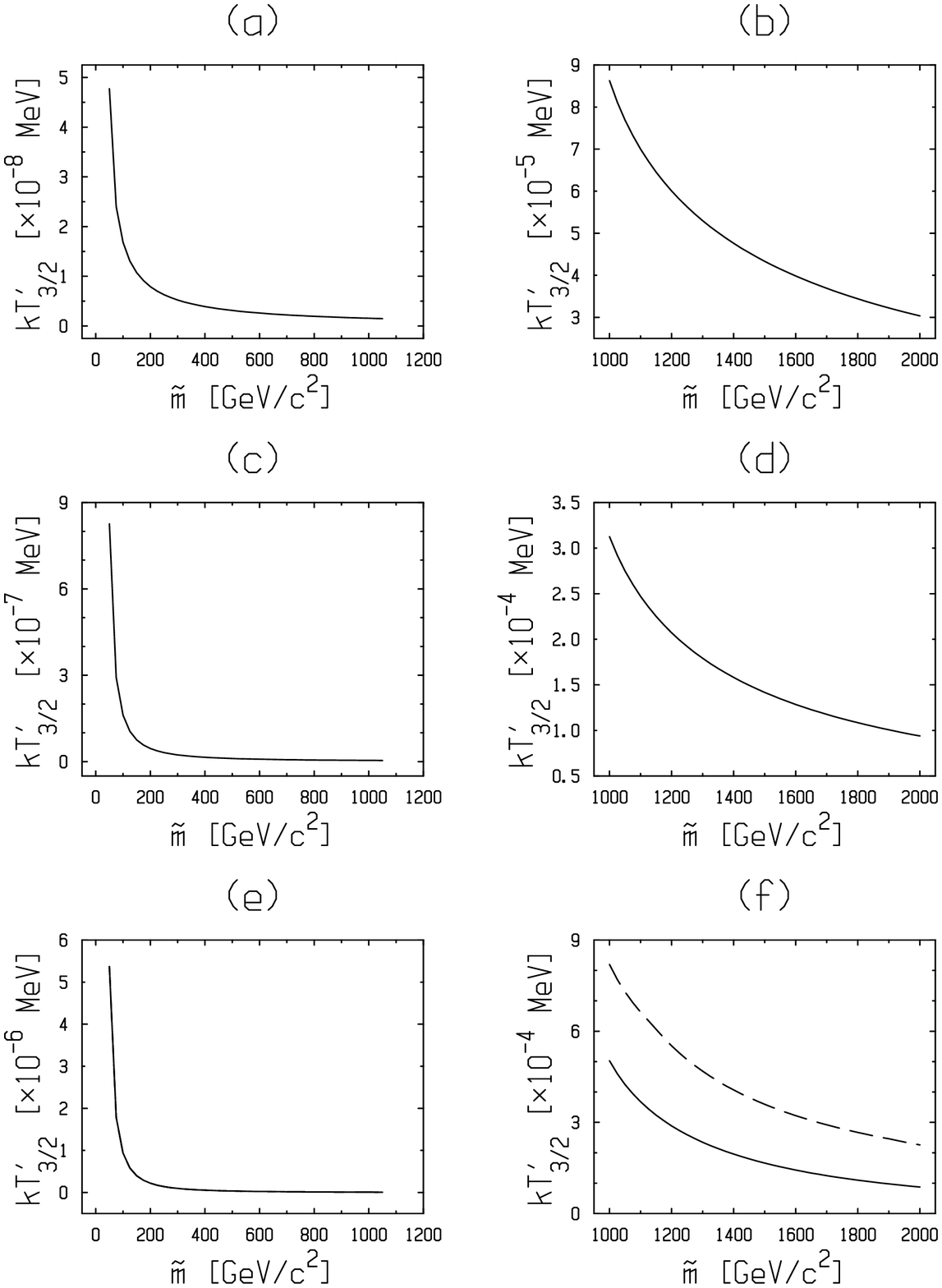}
\end{center}
\end{figure}
\clearpage
\newpage
\begin{figure}
\caption{Gravitino decay temperature $kT^\prime_{3/2}$ (in $MeV$)
as a function of the scalar superpartner mass $\tilde m$ (in $GeV/c^2$),
with $m_{3/2}=50GeV/c^2$ in Figures (a,c,e) and $m_{3/2}=1TeV/c^2$
in Figures (b,d,f). The temperature $T^\prime_{3/2}$ represented by a
plain line in Figures (a,b), (c,d) and (e,f) has been obtained
(via Eq.(\ref{Tcp2})) from the gravitino decay rates involving
the \rpv coupling constants $\l_{233}=0.06(\tilde m/100GeV)$,
$\l^\prime_{331}=0.48$ for $\tilde m=100GeV$ (see \cite{Ellis}
for the numerical dependence on $\tilde m$) and
$\l^{\prime \prime}_{213}=1.25$, respectively. The temperature
$T^\prime_{3/2}$ represented as a dashed line in Figure (f) has
been derived from the sum of gravitino decay rates involving
the following \rpv coupling constants, $\l^{\prime}_{132}=0.34$
for $\tilde m=100GeV$ (see \cite{Ellis} for the numerical
dependence on $\tilde m$), $\l^{\prime}_{211}=0.09(\tilde m
/100GeV)$, $\l^{\prime}_{223}=0.18(\tilde m
/100GeV)$, $\l^{\prime}_{311}=0.10(\tilde m/100GeV)$,
$\l_{121}=0.05(\tilde m/100GeV)$ and $\l_{233}=
0.06(\tilde m/100GeV)$.}
\label{astrodec}
\end{figure}
We observe in Fig.\ref{astrodec}(a) and Fig.\ref{astrodec}(b) a
decrease of the gravitino decay temperature $T^\prime_{3/2}$ when
the scalar superpartner becomes heavier (particularly in
Fig.\ref{astrodec}(a) for $m_{3/2} \sim 50GeV$). This is due to the
fact that $T^\prime_{3/2}$ is proportional (see Eq.(\ref{Tcp2})) to the
square root of the sum of gravitino decay rates $\Gamma(\tilde G
\stackrel{\l_{233}}{\to} \nu_\mu \tau \bar \tau)+\Gamma(\tilde G
\stackrel{\l_{323}}{\to} \nu_\tau \mu \bar \tau)$ which decreases as
the scalar superpartner mass increases. Nevertheless, this decrease
of $T^\prime_{3/2}$ when $\tilde m$ increases is attenuated by our choice
of taking $\l_{233}=0.06(\tilde m/100GeV)$, since $\Gamma(\tilde G
\stackrel{\l_{233}}{\to} \nu_\mu \tau \bar \tau) \propto \l^2_{233}$
and $\Gamma(\tilde G \stackrel{\l_{323}}{\to} \nu_\tau \mu \bar \tau)
\propto \l^2_{323}=\l^2_{233}$ (see Eq.(\ref{gamma})) so that $T^\prime_{3/2}
\propto \l_{233}$. We also remark by comparing Fig.\ref{astrodec}(a) and
Fig.\ref{astrodec}(b) that $T^\prime_{3/2}$ is getting higher if $m_{3/2}$
is enhanced. Once more, the explanation is related to the behavior
of the sum of gravitino decay rates $\Gamma(\tilde G
\stackrel{\l_{233}}{\to} \nu_\mu \tau \bar \tau)+\Gamma(\tilde G
\stackrel{\l_{323}}{\to} \nu_\tau \mu \bar \tau)$:  
those ones of course increase with $m_{3/2}$.
We conclude from Fig.\ref{astrodec}(a) and Fig.\ref{astrodec}(b)
that the gravitino decay temperature $T^\prime_{3/2}$ fulfills the
condition \ref{bound1} but not \ref{bound2}, in case one \rpv
coupling constant of type $\l_{ijk}$ is dominant. This conclusion
cannot be modified by the value of the factor
$g(T_d)^{1/4}/g(T_{3/2})^{1/2}$ of Eq.(\ref{Tcp2}) which is not likely
to be significantly large. We must also note that
only scalar superpartner masses smaller than ${\cal O}(TeV)$ have been
considered in Fig.\ref{astrodec}(a) and Fig.\ref{astrodec}(b). This
is to ensure that the ``hierarchy problem'', namely the problem of
natural coexistence of the electroweak symmetry breaking scale and
the scale of new physics underlying the Standard Model (grand
unification scale, string scale \dots), can be solved by the
presence of supersymmetry. Now, if the gravitino mass (which is also
the LSP mass), and thus the scalar superpartner masses, reach values
as large as $\sim 10^2 TeV$, the condition of Eq.(\ref{bound2}) can be
fulfilled, in case one \rpv coupling constant of type $\l_{ijk}$ is
dominant.

\vskip 0.4 cm {\bf $\bullet \ \lambda^\prime_{ijk}$ couplings} \vskip 0.4 cm 
Secondly, we suppose that one \rpv coupling constant of type
$\l^\prime_{ijk}$ is dominant compared with all the other \rpv
coupling constants. In this framework, the temperature
$T^\prime_{3/2}$ is then obtained (through Eq.(\ref{Tcp2})) from the whole
gravitino decay rate, which is the sum of the two gravitino decay rates
involving a $\l^\prime_{ijk}$ coupling constant, namely
$\Gamma(\tilde G \stackrel{\l^\prime_{ijk}}{\to} \nu_i d_j \bar d_k)$
and $\Gamma(\tilde G \stackrel{\l^\prime_{ijk}}{\to} e_i u_j \bar d_k)$
(see Section \ref{sec:rate}).
\\ In Fig.\ref{astrodec}(c) and Fig.\ref{astrodec}(d),
we show the temperature $T^\prime_{3/2}$ as a function of the scalar
superpartner mass $\tilde m$ for two gravitino masses, in the case
of a single dominant \rpv coupling constant of type $\l^\prime_{ijk}$.
The explanations on the calculation of $T^\prime_{3/2}$
concerning the numerical factors and the masses of the particles involved
in the gravitino decay process, which are given above for the case of
a dominant $\l_{ijk}$ coupling, still hold in the present case.
In order to optimize the temperature $T^\prime_{3/2}$, we have also supposed
here
that the single dominant \rpv coupling constant is $\l^\prime_{331}$,
which is one of the $\l^\prime_{ijk}$ coupling cons\-tants having the weakest
present low-energy constraint, namely $\l^\prime_{331}<0.48$ for
$m_{\tilde q}=100GeV$ \cite{Bhatt} (see \cite{Ellis} for the numerical
dependence of this limit on the squark mass), and that $\l^\prime_{331}$
is equal to its present maximum allowed value. The present low-energy
constraint on $\l^\prime_{331}$ has been derived by assuming that
$\l^\prime_{331}$ is the single dominant \rpv coupling constant
\cite{Ellis},
which is once more consistent with our assumptions. The gravitino decay
rate $\Gamma(\tilde G \stackrel{\l^\prime_{331}}{\to} \tau t \bar d)$
depends on the top quark mass that we have fixed at $m_{top}=174.2GeV$
\cite{top}.
\\ The dependence of the temperature $T^\prime_{3/2}$ on $\tilde m$ and
$m_{3/2}$ observed in Fig.\ref{astrodec}(c) and Fig.\ref{astrodec}(d)
is similar to that seen in Fig.\ref{astrodec}(a) and Fig.\ref{astrodec}(b),
which has been described above.
\\ It appears clearly in Fig.\ref{astrodec}(c) and Fig.\ref{astrodec}(d)
that the gravitino decay temperature $T^\prime_{3/2}$ res\-pects the
constraint of Eq.(\ref{bound1}) but not of Eq.(\ref{bound2}), in case
one \rpv coupling constant of type $\l^\prime_{ijk}$ is dominant. It must
be mentioned at this level that, as before, scalar superpartner masses
larger than ${\cal O}(TeV)$ have not been considered in order to preserve the
natural coexistence of electroweak symmetry breaking and new
physics scales, guaranteed by supersymmetric theories. We have found
that for gravitino masses greater than $\sim 10^2 TeV$, the constraint
\ref{bound2} can be respected in the case of a dominant $\l^\prime_{ijk}$
coupling.

\begin{figure}[t]
\centerline{\psfig{figure=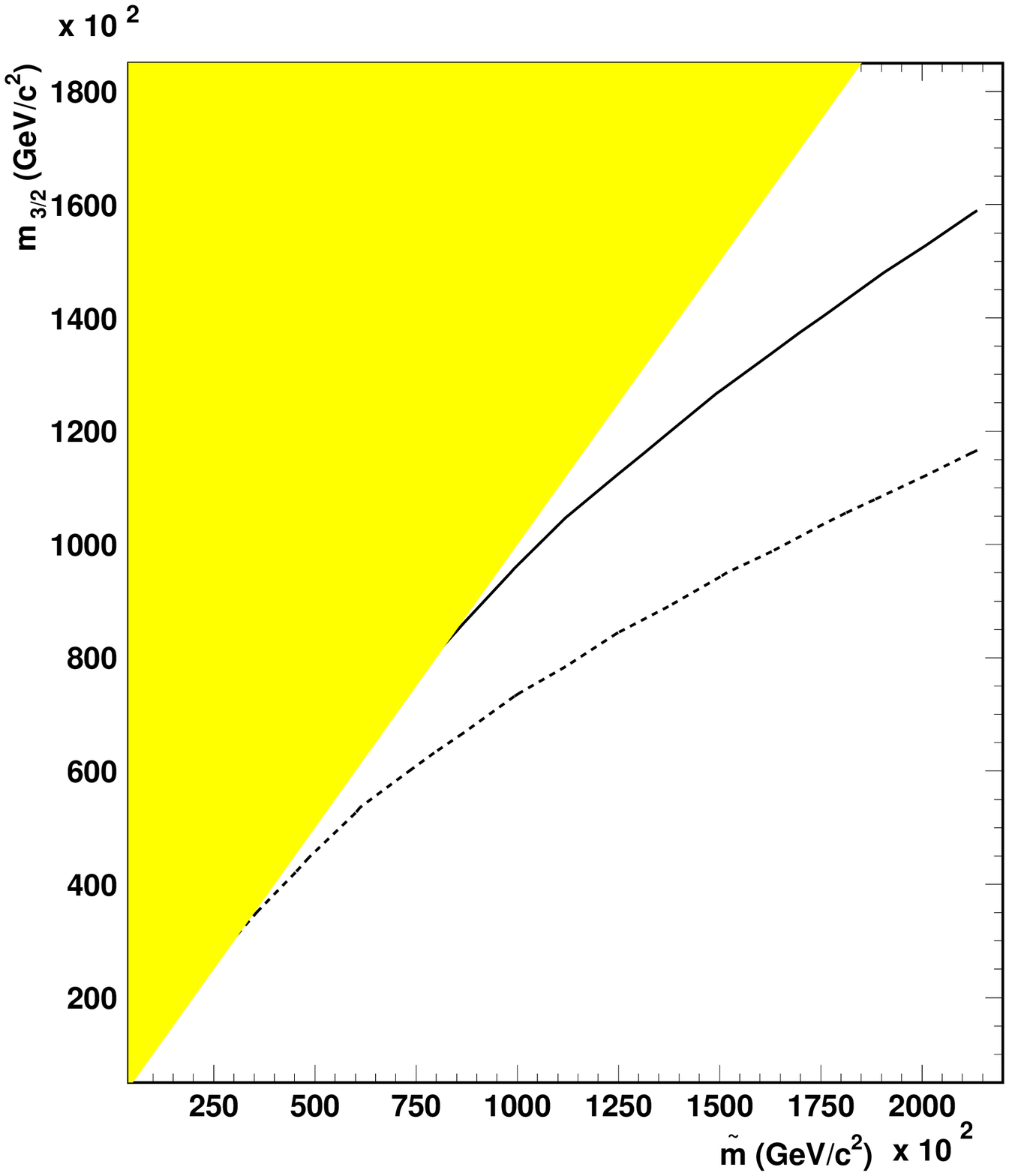,height=10.cm,width=10.cm}}
\caption{Domains of the $m_{3/2}(GeV/c^2)$-$\tilde m(GeV/c^2)$ plane
(gravitino versus superpartner mass) in which the gravitino decay
temperature is higher than the nucleosynthesis one, namely
$k T^\prime_{3/2} > 0.4MeV$. The region situated above the plain
(dashed) line corresponds to $k T^\prime_{3/2} > 0.4MeV$ in case the
do\-minant \rpv coupling constant is $\l^{\prime \prime}_{213}=1.25$
(the dominant \rpv coupling constants are $\l^{\prime}_{132}=0.34$
for $\tilde m=100GeV$, $\l^{\prime}_{211}=0.09(\tilde m
/100GeV)$, $\l^{\prime}_{223}=0.18(\tilde m
/100GeV)$, $\l^{\prime}_{311}=0.10(\tilde m/100GeV)$,
$\l_{121}=0.05(\tilde m/100GeV)$ and $\l_{233}=0.06(\tilde m/100GeV)$).
We note that the considered \rpv coupling cons\-tants are fixed
to the value of the perturbative limit ($\sqrt{4 \pi}$) in the mass domains
where their present bound exceeds this limit. Finally, the colored region
corresponds to the situation $m_{3/2}>\tilde m$ which must be considered
within a scenario where the gravitino is not the LSP.}
\label{niv}
\end{figure}

\vskip 0.4 cm {\bf $\bullet \ \lambda^{\prime \prime}_{ijk}$
couplings}. \vskip 0.4 cm
For the case of a single dominant \rpv coupling constant of type
$\l^{\prime \prime}_{ijk}$, the temperature $T^\prime_{3/2}$ is
deduced from the gravitino decay rate
involving $\l^{\prime \prime}_{ijk}$ coupling constants
$\Gamma(\tilde G \stackrel{\l^{\prime \prime}_{ijk}}{\to} u_i d_j d_k)$
(see Eq.(\ref{gammapp})).
\\ In Fig.\ref{astrodec}(e) and Fig.\ref{astrodec}(f) (plain line),
we have represented $T^\prime_{3/2}$ as a function of $\tilde m$ for two values
of $m_{3/2}$, in the case of a single dominant $\l^{\prime \prime}_{ijk}$
coupling constant. The explanations given above on the computation of
$T^\prime_{3/2}$
concerning the numerical factors, the particle masses and the anti-symmetry
of
the $\l_{ijk}$ coupling constants, still applies in the present case since
the $\l^{\prime \prime}_{ijk}$ coupling constants are anti-symmetric in the
$j$ and $k$ generation indices.
We have also supposed here that the single dominant \rpv coupling constant
is
$\l^{\prime \prime}_{213}$, which is one of the $\l^{\prime \prime}_{ijk}$
coupling constants having the weakest present bound, namely
$\l^{\prime \prime}_{213}<1.25$ for a supersymmetry breaking scale of
$M_{SUSY} \approx 1TeV$ \cite{Bra,Sher}, and that $\l^{\prime \prime}_{213}$
is equal to its present limit. Let us make a few remarks on the present
bound on $\l^{\prime \prime}_{213}$. This constraint is not, as the previous
ones, a low-energy experimental bound obtained at colliders: it originates
from
the requirement that $\l^{\prime \prime}_{213}$ remains in the perturbative
domain ($\l^{\prime \prime 2}_{213}/4 \pi<1$) up to the scale of gauge group
unification. We also mention that this constraint depends weakly on the
value
of $M_{SUSY}$, and that it has been obtained under the hypothesis that
$\l^{\prime \prime}_{213}$ is the dominant \rpv coupling constant.
\\ We see in Fig.\ref{astrodec}(e) and Fig.\ref{astrodec}(f)
(plain line) that the temperature $T^\prime_{3/2}$ fulfills the
condition \ref{bound1} but not \ref{bound2}.
\\ In Fig.\ref{niv} (plain line), we show the values of the gravitino and
superpartner masses for which $k T^\prime_{3/2} > 0.4MeV$. These values have
been
obtained by assuming that the dominant \rpv coupling constant is
$\l^{\prime \prime}_{213}$ and is equal to its present limit.
We see in Fig.\ref{niv} (plain line) that as the superpartner mass
increases,
larger gravitino masses are needed to have $k T^\prime_{3/2} > 0.4MeV$. The
reasons
are that $T^\prime_{3/2}$ increases with the gravitino mass but is suppressed if
the superpartner is getting heavier (see previous figures).
We can however conclude from Fig.\ref{niv} (plain line) that, typically,
gravitino (and scalar superpartner)
masses larger than $\sim 80 TeV$ are needed so that the condition of
Eq.(\ref{bound2}) can be fulfilled, in the case where $\l^{\prime
\prime}_{213}$
is the dominant coupling constant.

Finally, we consider the situation in which a maximum number 
of \rpv coupling constants having the weakest present bounds
are simultaneously dominant (and the considered \rpv coupling
constants are equal to their present limit, as before).
Based on the strongest constraints on the products of \rpv coupling
constants and on the review \cite{Bhatt} of the present limits
on the single \rpv coupling constants, we have found that this
optimistic situation corresponds to the case where the simultaneously
dominant \rpv coupling constants are,
\begin{displaymath}
\l^{\prime}_{132},\ \l^{\prime}_{211},\ \l^{\prime}_{223}, 
\ \l^{\prime}_{311},\ \l_{121} \ \mbox{and} \ \l_{233},
\end{displaymath}
which have the present bounds
$\l^{\prime}_{132}<0.34$ for $m_{\tilde q}=100GeV$ (and for instance
$\l^{\prime}_{132} \stackrel{<}{\sim} 1.2$ for $m_{\tilde q}=1TeV$) 
\cite{Bhatt,Ellis},
$\l^{\prime}_{211}<0.09(m_{\tilde d_R}$ $/100GeV)$ \cite{Barger},
$\l^{\prime}_{223}<0.18(m_{\tilde b_R}/100GeV)$ \cite{Deba},
$\l^{\prime}_{311}<0.10(m_{\tilde d_R}/100GeV)$ \cite{Bhatt,Deba},
$\l_{121}<0.05(m_{\tilde e_R} /100GeV)$ \cite{Drein,Barger}
and $\l_{233}<0.06(m_{\tilde \tau_R}/100GeV)$
\cite{Drein,Barger}. At this stage, an important
point to note is that the $\l^{\prime}_{ijk}$ and
$\l^{\prime \prime}_{ijk}$ couplings cannot be simultaneously
dominant, since the experimental cons\-traints on
the proton decay rate force any product $\l^{\prime}_{ijk}
\l^{\prime \prime}_{i'j'k'}$ to be smaller than $10^{-9}$, in a
conservative way and for squark masses below $1 TeV$. This result
has been obtained in \cite{Smirnov} by calculating the proton
decay rate at one loop level. In contrast, we have checked here
that no strong constraint exists on any product taken among the
\rpv coupling constants $\l^{\prime}_{132}$, $\l^{\prime}_{211}$, 
$\l^{\prime}_{223}$,
$\l^{\prime}_{311}$, $\l_{121}$ and $\l_{233}$. We also note that in
the optimistic situation described above, no other \rpv coupling
constant of type $\l^{\prime}_{22k}$ ($\l^{\prime}_{31k}$) can be supposed
simultaneously dominant, since the present constraint on
$\l^{\prime}_{223}$ ($\l^{\prime}_{311}$) has been obtained under the
hypothesis that $\l^{\prime}_{223}$ ($\l^{\prime}_{311}$) is the only
dominant coupling constant among the $\l^{\prime}_{22k}$
($\l^{\prime}_{31k}$) coupling constants \cite{Deba}. This
interpretation corresponds admittedly to a strong version of the
single coupling constant dominance hypothesis implying that the
contributions from the subdominant coupling constants do not interfere
destructively in the observable used to infer the bound.
Similarly, the present bound on $\l^{\prime}_{132}$
\cite{Bhatt,Ellis} (a coupling constant of type $\l^{\prime}_{32k}$
\cite{Drein,Aga}) has been derived by assuming that it dominates over all the
$\l^{\prime}_{1jk}$
($\l^{\prime}_{i1k}$ and $\l^{\prime}_{i2k}$) coupling constants, so that no
additional \rpv coupling constant of type $\l^{\prime}_{1jk}$
($\l^{\prime}_{32k}$)
can be assumed simultaneously dominant. Once again, if we use the present bound on
$\l^{\prime}_{211}$, all the other $\l^{\prime}_{21k}$ and $\l^{\prime}_{11k}$ coupling 
constants must neglected \cite{Barger}, so that no additional \rpv coupling constant 
of type $\l^{\prime}_{21k}$ can be considered simultaneously.
$\l^{\prime}_{231}$ also cannot be simultaneously dominant, for another reason
which is the existence of the following constraint,
\begin{displaymath} 
\l^{\prime}_{i31} \l^{\prime}_{132}  
\ < \ 4.7 \ 10^{-5} \ (m_{\tilde b_L}/100GeV)^2 \ \cite{Pro}.
\end{displaymath}
Finally, the remaining $\l^{\prime}$ coupling constants, $\l^{\prime}_{23k}$ ($k=2,3$) 
and $\l^{\prime}_{33k}$ ($k=1,2,3$), cannot be additional dominating couplings,
because the present bounds on those have been obtained by assuming the dominance 
over all the $\l^{\prime}_{2jk}$ and $\l^{\prime}_{3jk}$ couplings, 
respectively \cite{Bhatt,Ellis}.
Concerning the $\l_{ijk}$ coupling constants, the constraint on $\l_{121}$
($\l_{233}$) has been obtained under the assumption that $\l_{121}$
($\l_{233}$) dominates among the $\l_{12k}$ and $\l^{\prime}_{11k}$
($\l_{13k}$ and $\l_{23k}$) coupling constants \cite{Barger}.
Therefore, in the optimistic framework described above, while the hypothesis
that $\l_{121}$ and $\l_{233}$ simultaneously dominate remains consistent
with the assumptions adopted in \cite{Barger}, no additional \rpv coupling
constant of type $\l_{ijk}$ can be considered simultaneously.
\\ In Fig.\ref{astrodec}(f) (dashed line),
we have represented $T^\prime_{3/2}$ as a function of $\tilde m$,
in the optimistic situation described above. In this situation,
the temperature $T^\prime_{3/2}$ is derived from the sum of the gravitino
decay rates involving the coupling constants $\l^{\prime}_{132}$,
$\l^{\prime}_{211}$, $\l^{\prime}_{223}$,  
$\l^{\prime}_{311}$, $\l_{121}$, $\l_{211}$, $\l_{233}$ and $\l_{323}$.
\\ We see in Fig.\ref{astrodec}(f) (dashed line) that
the temperature $T^\prime_{3/2}$ fulfills the condition \ref{bound1} but not
\ref{bound2}. In fact, gravitino masses larger than $\sim 30 TeV$ are needed
so that the condition \ref{bound2} can be fulfilled. This can be observed
on Fig.\ref{niv} (dashed line) in which have been represented the values of
the
gravitino and superpartner masses for which $k T^\prime_{3/2} > 0.4MeV$. Those
values
have been derived within the optimistic scenario described above.
In this optimistic scenario where several \rpv coupling constants are
simultaneously
present, the gravitino masses corresponding to $k T^\prime_{3/2} > 0.4MeV$ are
smaller
than in the cases of a single domi\-nant \rpv coupling constant, since as we
have already
seen $T^\prime_{3/2}$ is typically proportional to the \rpv coupling constants
and decreases
with the gravitino mass.
It is also interes\-ting to note that in this optimistic 
scenario, for a value of the ratio $g(T_d)^{1/4}/g(T_{3/2})^{1/2}$
(see Eq.(\ref{Tcp2})) of order $50$, the condition \ref{bound2}
is fulfilled for LSP gravitino masses of ${\cal O}(TeV)$ so that the
constraint $\tilde m \stackrel{<}{\sim} {\cal O}(TeV)$ due to the hierarchy
problem can be respected. However, for $g(kT_{3/2} \approx 5 \ 10^{-3}MeV)
\approx 5$ (see Eq.(\ref{Tcinc})), corresponding to the value of the
$SU(3)_C \times SU(2)_L \times U(1)_Y$ model, $g(T_d)^{1/4}/
g(kT_{3/2} \approx 5 \ 10^{-3}MeV)^{1/2} \approx 50$ leads to
$g(T_d) \approx 10^8$ which is not realistic within the present models.

In summary, if the gravitino mass, and thus the scalar superpartner masses, 
do not exceed the $TeV$ scale by at least one order of magnitude,
the gravitino decay temperature $T^\prime_{3/2}$ fulfills well the constraint
of
Eq.(\ref{bound1}) but not of Eq.(\ref{bound2}), within the scenario of an
unstable
LSP gravitino decaying via $\l$, $\l^\prime$ or $\l^{\prime \prime}$
interactions.
Hence, this scenario does not appear to be a natural solution to the
cosmological
gravitino problem.

\section{Conclusion}

Along with the existence of the cosmic microwave background, big-bang
nucleosynthesis
is one of the most important predictions of the big-bang cosmology.
Furthermore, if
one assumes that the light nuclei (atomic number less than $7$) have
effectively been
produced through the big-bang nucleosynthesis, one finds that the
theoretical
predictions on the abundances of these light nuclei are in good agreements
with
the observational data \cite{Walker}. Now, as we have seen above, if one
believes that
the light elements have been synthesized through the standard big-bang
nucleosynthesis,
the gravitinos must have decayed before the nucleosynthesis epoch.
Therefore,
since we have found that this cannot happen in the scenario characterized
by an unstable LSP gravitino having a decay channel which involves trilinear
\rpv
interactions, this scenario does not seem to provide a realistic solution
to the
large relic abundance of the gravitino. It must be specified that this
conclusion
has been obtained by assuming that the gravitino and thus the scalar
superpartner
masses do not exceed the $TeV$ scale so that supersymmetry can solve
the
so-called hierarchy problem, which is one of the strongest motivations for
the
existence of supersymmetry. We also mention that this final conclusion is
based on
the present constraints on the trilinear \rpv couplings \cite{Bhatt}
and on the assumption that the ratio $g(T_d)^{1/4}/g(T_{3/2})^{1/2}$
takes reasonable values.

The authors of \cite{Taka} have found that the lifetime of
the partial two-body gravitino decay $\tilde G \to \gamma + \nu$, 
initiated by the bilinear \rpv interactions with a coupling constant 
yielding contributions to the neutrino mass matrix which match the
observed atmospheric neutrino anomaly, reads as: 
\begin{eqnarray}
\tau_{3/2} (\tilde G \to \gamma + \nu) 
\approx 8.3 \ 10^{26} \ 
\bigg ( {1 GeV \over \mm} \bigg )^3 \ sec,
\label{lifetime}
\end{eqnarray}
where the neutrino-photino mixing matrix element was set at the value
appropriate to a photino with a mass of $80 GeV$.
If one assumes tentatively that the relation in Eq.(\ref{lifetime}) 
continues to hold in order of magnitude for the heavier gauginos case, 
then the condition that LSP gravitinos decay radiatively before the
nucleosynthesis epoch, namely $\tau_{3/2} \stackrel{<}{\sim} 10^2 sec$, 
requires the lower mass bound $\mm \stackrel{>}{\sim} 2 \ 10^8 GeV$
which conflicts with the constraint from the hierarchy problem. 
On the other hand, the case of light LSP gravitinos, say with mass 
$m_{3/2} < 1 GeV$, leads within the bilinear \rpv option to very long
lived gravitinos which could constitute a valid candidate for the dark
matter \cite{Taka}. 
\\ Within the \rpv trilinear interactions
option, by invoking the bounds on coupling constants 
rather than the values which match the observations involving the 
neutrinos mass matrix, we obtain values of the 
gravitino lifetime which are significantly shorter 
than those obtained within the bilinear interactions option
by Takayama and Yamaguchi \cite{Taka}. As a function of the relevant 
ratio of squared masses of the gravitino and exchanged sfermions,
$z = m_{3/2} ^2 /\tilde m ^2 $, 
the predicted partial gravitino lifetime, associated to the decay 
into a given flavor configuration of three light fermions 
final state, decreases monotonically in the interval $0<z<1$ as,
\begin{eqnarray} 
\tau_{3/2} (\tilde G \to f_i f_j f_k)
\approx 10^9 - 10^{11} \ \bigg ( {1\over \hat \l_{ijk}^2} \bigg )
\bigg ( {1 TeV \over \mm} \bigg )^3  \ sec,
\label{lifetimeII}
\end{eqnarray} 
where $\hat \l$ stands for any one of the trilinear coupling
constants. For light gravitinos of mass 
$m_{3/2} = {\cal O}(GeV)$ and $\hat \l_{ijk} = {\cal O}(1)$, 
we see from Eq.(\ref{lifetimeII}) that
the gravitino lifetime exceeds by a few order of magnitudes 
the age of the universe $t_0 \simeq 3.2 \ 10^{17} \ sec$.

We have found that the present bounds on the trilinear \rpv
couplings allow easily the gravitinos to decay before the present epoch.
In consequence, the scenario of an un\-stable LSP gravitino having a decay
channel
which involves trilinear \rpv interactions could pos\-sibly solve the
cosmological
gravitino problem in the context of an inflatio\-nary model \cite{Moroi}.
The reason is that within such a framework, the gravitino would not have
necessa\-rily to decay before the nucleosynthesis epoch,
since in inflationary models small gravitino number densities can be
predicted (for small enough gravitino number densities, 
the gra\-vitino decays do not upset the nucleosynthesis).
Indeed, if the universe has experienced inflationary expansion, the
primordial
gra\-vitino abundance would have been completely diluted by the exponential
expansion. Then, after the universe would have been reheated, gravitinos
would have been regenerated by scattering processes off the thermal
radiation, resulting in a secondary gravitino number density proportional
to the reheating tempe\-rature, which is a parameter of the model.
\newline
\newline
{\bf \Large Acknowledgments}
\newline
\newline
We are grateful to R.~Grimm and F.~Takayama for helpful discussions.

\newpage

\appendix

\renewcommand{\thesubsection}{A.\arabic{subsection}}
\renewcommand{\theequation}{A.\arabic{equation}}
\setcounter{subsection}{0}
\setcounter{equation}{0}

\section{Definitions of the functions involved
in the gravitino decay rates}
\label{definitions}

The mass functions $\Omega(m)$, $\Xi(m_1,m_2)$, $\Delta(m_1,m_2)$
and $\Sigma(m_1,m_2)$ used in Section \ref{sec:rate} are defined
in the following,
\begin{eqnarray}
\Omega(m) & = & {1 \over 48}
\bigg ( 60 m^6-162 m^4 \mm^2+140 m^2 \mm^4-37 \mm^6 \cr &&
+12 (m^2-\mm^2)^3 (5 {m^2 \over \mm^2}-1) \log(1-{\mm^2 \over m^2}) \bigg ),
\label{fcta}
\end{eqnarray}
\begin{eqnarray}
\Xi(m_1,m_2)&=&{1 \over 48} \bigg (
84 m_1^2 m_2^2 \mm^2 + 7 \mm^6
+ m_1^2 (-12m_1^4+24m_2^4+42m_1^2 \mm^2 -40\mm^4) \cr
&&+ m_2^2 (-12m_2^4+24m_1^4+42m_2^2 \mm^2 -40\mm^4)
+ 12m_1^2 ({2m_2^2 m_1^4-m_1^6 \over \mm^2} +4m_1^4 \cr
&&-4m_2^2 \mm^2
-5m_1^2 \mm^2+2\mm^4+2 m_1^2 m_2^2) \log (1-{\mm^2 \over m_1^2})+12m_2^2 \cr
&& ({2m_1^2 m_2^4-m_2^6 \over \mm^2}+4m_2^4-4m_1^2 \mm^2
-5m_2^2 \mm^2+2\mm^4+2 m_1^2 m_2^2) \cr
&& \log (1-{\mm^2 \over m_2^2})
\bigg ),
\label{fctb}
\end{eqnarray}
\begin{eqnarray}
\Delta(m_1,m_2)= - m_1^2 m_2^2
({m_1^2 m_2^2 \over 2 \mm^2}+m_1^2 +m_2^2 -\mm^2),
\label{fctc}
\end{eqnarray}
\begin{eqnarray}
\Sigma(m_1,m_2)= \Delta(m_1,m_2)
\log (1-{\mm^2 \over m_1^2}) \log (1+{m_1^2 \over m_2^2}-{\mm^2 \over
m_2^2}),
\label{fctd}
\end{eqnarray}
$m_{3/2}$ being the gravitino mass.

\renewcommand{\thesubsection}{B.\arabic{subsection}}
\renewcommand{\theequation}{B.\arabic{equation}}
\setcounter{subsection}{0}
\setcounter{equation}{0}

\section{Formulas for the spin summed amplitudes
of the gravitino decay reactions}
\label{formulas}



The full analytical result of the scared amplitude summed over the
spins for the gravitino decay reaction
$\tilde G \stackrel{\l_{ijk}}{\to} \nu_i e_j \bar e_k$ is the sum
of the following scared amplitude and interference terms
\footnote{The projective sums for a massive spin-$3/2$ field 
are required in order to calculate the spin averaged 
gravitino decay amplitudes. These projective sums are
available in the literature \cite{Krauss,Moroi,psum}.},
\begin{eqnarray}
\vert M_a \vert^2 &=& {1 \over 3} {\l_{ijk}^2 \over M_\star^2
(m^2_{jk}-\aa^2)^2}
(\mm^2-\mjk^2+\mi^2) (\mjk^2-\mj^2-\mk^2) \cr &&
\bigg ( {(\mm^2+\mjk^2-\mi^2)^2 \over 4 \mm^2}-\mjk^2 \bigg ),
\label{amp1}
\end{eqnarray}
\begin{eqnarray}
\vert M_{b} \vert^2 &=& {1 \over 3} {\l_{ijk}^2 \over M_\star^2
(m^2_{ik}-\bb^2)^2}
(\mm^2-\mik^2+\mj^2) (\mik^2-\mi^2-\mk^2)  \cr &&
\bigg ( {(\mm^2+\mik^2-\mj^2)^2 \over 4 \mm^2}-\mik^2 \bigg ),
\label{amp2}
\end{eqnarray}
\begin{eqnarray}
\vert M_{c} \vert^2 &=& {1 \over 3} {\l_{ijk}^2 \over M_\star^2
(m^2_{ij}-\cc^2)^2}
(\mm^2-\mij^2+\mk^2) (\mij^2-\mi^2-\mj^2) \cr &&
\bigg ( {(\mm^2+\mij^2-\mk^2)^2 \over 4 \mm^2}-\mij^2 \bigg ),
\label{amp3}
\end{eqnarray}
\begin{eqnarray}
2 Re(M_a M^\dagger_{b})&=&{1 \over 3} {\l_{ijk}^2
\over M_\star^2 (m^2_{jk}-\aa^2) (m^2_{ik}-\bb^2)} \bigg [
(\mik^2 \mjk^2 - \mm^2 \mk^2 - \mi^2 \mj^2) \cr && \bigg (
(\mm^2+\mk^2-\mi^2-\mj^2)
-{1 \over 2 \mm^2}(\mm^2+\mjk^2-\mi^2) \cr && (\mm^2+\mik^2-\mj^2) \bigg )
+ {1 \over 2} (\mij^2 - \mi^2 - \mj^2) (\mjk^2 - \mj^2 - \mk^2) \cr &&
(\mik^2 - \mi^2 - \mk^2) - {\mi^2 \over 2} (\mjk^2 - \mj^2 - \mk^2)^2
- {\mj^2 \over 2} (\mik^2 - \mi^2 - \mk^2)^2 \cr &&
- {\mk^2 \over 2} (\mij^2 - \mi^2 - \mj^2)^2 + 2 \mi^2 \mj^2 \mk^2 \bigg ],
\label{amp12}
\end{eqnarray}
\begin{eqnarray}
2 Re(M_{b} M^\dagger_{c}) &=& {1 \over 3} {\l_{ijk}^2
\over M_\star^2 (m^2_{ik}-\bb^2) (m^2_{ij}-\cc^2)} \bigg [
(\mij^2 \mik^2 - \mm^2 \mi^2 - \mj^2 \mk^2) \cr && \bigg (
(\mm^2+\mi^2-\mj^2-\mk^2)
-{1 \over 2 \mm^2}(\mm^2+\mik^2-\mj^2) \cr && (\mm^2+\mij^2-\mk^2) \bigg )
+ {1 \over 2} (\mij^2 - \mi^2 - \mj^2) (\mjk^2 - \mj^2 - \mk^2) \cr &&
(\mik^2 - \mi^2 - \mk^2) - {\mi^2 \over 2} (\mjk^2 - \mj^2 - \mk^2)^2
- {\mj^2 \over 2} (\mik^2 - \mi^2 - \mk^2)^2 \cr &&
- {\mk^2 \over 2} (\mij^2 - \mi^2 - \mj^2)^2 + 2 \mi^2 \mj^2 \mk^2 \bigg ],
\label{amp23}
\end{eqnarray}
\begin{eqnarray}
2 Re(M_a M^\dagger_{c}) &=& {1 \over 3} {\l_{ijk}^2
\over M_\star^2 (m^2_{jk}-\aa^2) (m^2_{ij}-\cc^2)} \bigg [
(\mij^2 \mjk^2 - \mm^2 \mj^2 - \mi^2 \mk^2) \cr && \bigg (
(\mm^2+\mj^2-\mi^2-\mk^2)
-{1 \over 2 \mm^2}(\mm^2+\mjk^2-\mi^2) \cr && (\mm^2+\mij^2-\mk^2) \bigg )
+ {1 \over 2} (\mij^2 - \mi^2 - \mj^2) (\mjk^2 - \mj^2 - \mk^2) \cr &&
(\mik^2 - \mi^2 - \mk^2) - {\mi^2 \over 2} (\mjk^2 - \mj^2 - \mk^2)^2
- {\mj^2 \over 2} (\mik^2 - \mi^2 - \mk^2)^2 \cr &&
- {\mk^2 \over 2} (\mij^2 - \mi^2 - \mj^2)^2 + 2 \mi^2 \mj^2 \mk^2 \bigg ],
\label{amp13}
\end{eqnarray}
where $M_a$, $M_b$ and $M_{3/2}$ correspond to the amplitudes of the three
contributions represented in Fig.\ref{grdec1} and marked by the same
letters $a$, $b$ and $c$. 
In Eq.(\ref{amp1}) to Eq.(\ref{amp13}), we have used the same
notations as in Section \ref{sec:rate} and the quantities $\mij$, $\mjk$
and $\mik$ are defined by,
\begin{eqnarray}
m^2_{ij} & = & (p(\nu_i)+p(e_j))^2 \cr
m^2_{jk} & = & (p(  e_j)+p(e_k))^2 \cr
m^2_{ik} & = & (p(\nu_i)+p(e_k))^2,
\label{mcar}
\end{eqnarray}
$p(\nu_i)$, $p(e_j)$ and $p(e_k)$ being respectively the quadri-momenta
of the particles $\nu_i$, $e_j$ and $e_k$. The formula for the spin summed
amplitude of the gravitino decay process $\tilde G \stackrel{\l_{ijk}}{\to}
\nu_i e_j \bar e_k$, which is given in Eq.(\ref{amp1})-Eq.(\ref{amp13}),
is equal to that of the charge conjugated process
$\tilde G \stackrel{\l_{ijk}}{\to} \bar \nu_i \bar e_j e_k$.

The full analytical result of the scared amplitude summed over the
spins for the gra\-vitino decay reaction $\tilde G
\stackrel{\l^\prime_{ijk}}{\to} \nu_i d_j \bar d_k$ ($\tilde G
\stackrel{\l^\prime_{ijk}}{\to} e_i u_j \bar d_k$) is equal to that of
the reaction $\tilde G \stackrel{\l_{ijk}}{\to} \nu_i e_j \bar e_k$ given
above, simply by changing $\l_{ijk}$ to $\l^\prime_{ijk}$, adding an extra
multiplicative color factor of $N_c=3$ and replacing the masses as follows:
$\aa \to \aaps \ (\aapt)$, $\bb \to \bbps \ (\bbpt)$, $\cc \to \ccps \
(\ccpt)$,
$\mi \to \mips \ (\mipt)$, $\mj \to \mjps \ (\mjpt)$ and
$\mk \to \mkps \ (\mkpt)$. The definitions of the quantities $\mij$, $\mjk$
and $\mik$ must also be changed into,
\begin{eqnarray}
m^2_{ij} & = & (p(\nu_i)+p(d_j))^2 \cr
m^2_{jk} & = & (p(  d_j)+p(d_k))^2 \cr
m^2_{ik} & = & (p(\nu_i)+p(d_k))^2,
\label{mcarp1}
\end{eqnarray}
for the spin summed amplitude of the reaction $\tilde G
\stackrel{\l^\prime_{ijk}}{\to} \nu_i d_j \bar d_k$, and into,
\begin{eqnarray}
m^2_{ij} & = & (p(e_i)+p(u_j))^2 \cr
m^2_{jk} & = & (p(u_j)+p(d_k))^2 \cr
m^2_{ik} & = & (p(e_i)+p(d_k))^2,
\label{mcarp2}
\end{eqnarray}
for the spin summed amplitude of the reaction $\tilde G
\stackrel{\l^\prime_{ijk}}{\to} e_i u_j \bar d_k$. The same remark as
above holds for the charge conjugated process.

The full analytical result of the scared amplitude summed over the
spins for the gra\-vitino decay reaction
$\tilde G \stackrel{\l^{\prime \prime}_{ijk}}{\to} u_i d_j d_k$ is
the sum of the following scared amplitude and interference terms,
\begin{eqnarray}
\vert M_a \vert^2 &=& { N_c ! \over 3}
{\l^{\prime \prime 2}_{ijk} \over M_\star^2 (m^2_{jk}-\aapp^2)^2}
(\mm^2-\mjk^2+\mipp^2) (\mjk^2-\mjpp^2-\mkpp^2) \cr &&
\bigg ( {(\mm^2+\mjk^2-\mipp^2)^2 \over 4 \mm^2}-\mjk^2 \bigg ),
\label{amp1pp}
\end{eqnarray}
\begin{eqnarray}
\vert M_{b} \vert^2 &=& {4 N_c ! \over 3} {\l^{\prime \prime 2}_{ijk} \over
M_\star^2
(m^2_{ik}-\bbpp^2)^2}
(\mm^2-\mik^2+\mjpp^2) (\mik^2-\mipp^2-\mkpp^2)  \cr &&
\bigg ( {(\mm^2+\mik^2-\mjpp^2)^2 \over 4 \mm^2}-\mik^2 \bigg ),
\label{amp2pp}
\end{eqnarray}
\begin{eqnarray}
2 Re(M_a M^\dagger_{b})&=&{2 N_c ! \over 3} {\l^{\prime \prime 2}_{ijk}
\over M_\star^2 (m^2_{jk}-\aapp^2) (m^2_{ik}-\bbpp^2)} \bigg [
(\mik^2 \mjk^2 - \mm^2 \mkpp^2 - \mipp^2 \mjpp^2) \cr &&
\bigg ( (\mm^2+\mkpp^2-\mipp^2-\mjpp^2)
-{1 \over 2 \mm^2}(\mm^2+\mjk^2-\mipp^2) \cr && (\mm^2+\mik^2-\mjpp^2) 
\bigg )
+ {1 \over 2} (\mij^2 - \mi^2 - \mj^2) (\mjk^2 - \mj^2 - \mk^2) \cr &&
(\mik^2 - \mi^2 - \mk^2) - {\mi^2 \over 2} (\mjk^2 - \mj^2 - \mk^2)^2
- {\mj^2 \over 2} (\mik^2 - \mi^2 - \mk^2)^2 \cr &&
- {\mk^2 \over 2} (\mij^2 - \mi^2 - \mj^2)^2 + 2 \mi^2 \mj^2 \mk^2 \bigg ],
\label{amp12pp}
\end{eqnarray}
where $M_a$ and $M_b$ are associated to the amplitudes of the two
contributions represented in Fig.\ref{grdec3} and marked by the same
letters $a$, $b$ and $c$. 
In Eq.(\ref{amp1pp}) to Eq.(\ref{amp12pp}), the notations of
Section \ref{sec:rate} have been used and the quantities $\mij$, $\mjk$
and $\mik$ are defined as follows,
\begin{eqnarray}
m^2_{ij} & = & (p(u_i)+p(d_j))^2 \cr
m^2_{jk} & = & (p(d_j)+p(d_k))^2 \cr
m^2_{ik} & = & (p(u_i)+p(d_k))^2.
\label{mcarpp}
\end{eqnarray}
The above remark on the charge conjugated process still holds.


\end{document}